# Security for Grid Services

Von Welch[1]    Frank Siebenlist[2]    Ian Foster[1,2]    John Bresnahan[2]
Karl Czajkowski[3]    Jarek Gawor[2]    Carl Kesselman[3]    Sam Meder[1]
Laura Pearlman[3]    Steven Tuecke[2]

[1] University of Chicago, Department of Computer Science
[2] Argonne National Laboratory, Mathematics and Computer Science Division
[3] University of Southern California, Information Sciences Institute

{welch, siebenlist, foster, bresnahan, gawor, meder, tuecke}@mcs.anl.gov
{karlcz, carl, laura}@isi.edu

**Abstract**

*Grid computing is concerned with the sharing and coordinated use of diverse resources in distributed "virtual organizations." The dynamic and multi-institutional nature of these environments introduces challenging security issues that demand new technical approaches. In particular, one must deal with diverse local mechanisms, support dynamic creation of services, and enable dynamic creation of trust domains. We describe how these issues are addressed in two generations of the Globus Toolkit®. First, we review the Globus Toolkit version 2 (GT2) approach; then, we describe new approaches developed to support the Globus Toolkit version 3 (GT3) implementation of the Open Grid Services Architecture, an initiative that is recasting Grid concepts within a service-oriented framework based on Web services. GT3's security implementation uses Web services security mechanisms for credential exchange and other purposes, and introduces a tight least-privilege model that avoids the need for any privileged network service.*

## 1   Introduction

The term "Grid" refers to systems and applications that integrate and manage resources and services distributed across multiple control domains [12]. Pioneered in an e-science context, Grid technologies are also generating interest in industry, as a result of their apparent relevance to commercial distributed-computing applications [14].

A common scenario within Grid computing involves the formation of dynamic "virtual organizations" (VOs) [16] comprising groups of individuals and associated resources and services united by a common purpose but not located within a single administrative domain. The need to support the integration and management of resources within VOs introduces challenging security issues [15]. For a variety of issues relating to certification, group membership, authorization, and the like, the relationships among participants in VOs represent an overlay with respect to the relationships existing between those participants and their parent organizations. This overlay exists both in terms of trust and with respect to the security mechanisms and policies in place at those parent organizations.

Grid computing research has produced security technologies based not on direct interorganizational trust relationships but rather on the use of the VO as a bridge among the entities participating in a particular community or function. The results of this research have been incorporated into a widely used software system called the Globus Toolkit®[1] (GT) [4] that uses public key technologies to address issues of single sign-on, delegation [17], and identity mapping, while supporting standardized APIs such as GSS-API [23]. The Grid Security Infrastructure (GSI) is the name given to the portion of the Globus Toolkit that implements security functionality.

The recent definition of the Open Grid Services Infrastructure specification and other elements of the Open Grid Services Architecture (OGSA) [14] within the Global Grid Forum introduces new challenges and opportunities for Grid security. In particular, integration with Web services and hosting environment technologies introduces opportunities to leverage emerging security standards and technologies such as the Security Assertion Markup Language (SAML) [32] and Web services security.

Integration of GSI with OGSA enables the use of Web services techniques to express and publish policy [2], allowing applications to determine automatically what security policies and mechanisms are required of them. Implementing security in the form of OGSA services allows those services to be used as needed by applications to meet these requirements. Advanced hosting environments enable security functionality to be implemented outside of the application, simplifying development.

The remainder of this article is organized as follows. In Sections 2 and 3, we review Grid security challenges and GT2 security mechanisms. In Sections 4 and 5, we introduce our view of an OGSA security model and our GT3 implementation. We conclude in Section 6 with a brief discussion of future work.

## 2  Grid Security Challenges

Security requirements within the Grid environment are driven by the need to support scalable, dynamic, distributed virtual organizations (VOs) [16]—collections of diverse and distributed individuals that seek to share and use diverse resources in a coordinated fashion. From a security perspective, a key attribute of VOs is that participants and resources are governed by the rules and policies of the classical organizations of which they are members. Furthermore, while some VOs, such as multiyear scientific collaborations, may be large and long-lived (in which case explicit negotiations with resource providers are acceptable), others will be short-lived—created, perhaps, to support a single task, for example, two individuals sharing documents and data as they write a proposal—in which case overheads associated with VO creation and operation have to be small.

A fundamental requirement is thus to enable VO access to resources that exist within classical organizations and that, from the perspective of those classical organizations, have policies in place that speak only about local users. This VO access must be established and coordinated only through *binary* trust relationships that exist between (a) the local user and their organization and (b) the VO and the user. We cannot, in general, assume trust relationships between the classical organization and the VO or its external members.

Grid security mechanisms address these challenges by allowing a VO to be treated as a *policy domain overlay* as shown in Figure 1. Multiple resources or organizations outsource certain policy control(s) to a third party, the VO, which coordinates the outsourced policy in a consistent manner to allow for coordinated resource sharing and use.

Complicating Grid security is the fact that new services (i.e., resources) may be deployed and instantiated dynamically over a VO's lifetime. For example, a user may establish personal stateful interfaces to existing resources, or the VO itself may create directory services to keep track of VO participants. Like their static counterparts, these resources must be securely coordinated and must interact with other services.

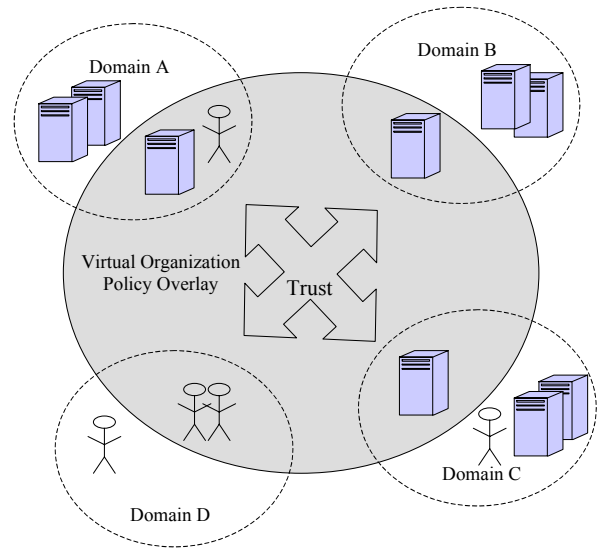

**Figure 1. A virtual organization policy domain overlay pulls together participants from disparate domains into a common trust domain.**

This combination of dynamic policy overlays and dynamically created entities drives the need for three key functions in a Grid security model.

1. *Multiple security mechanisms*. Organizations participating in a VO often have significant investment in existing security mechanisms and infrastructure. Grid security must interoperate with, rather than replace, those mechanisms.
2. *Dynamic creation of services*. Users must be able to create new services (e.g., "resources") dynamically without administrator intervention. These services must be coordinated and must interact securely with other services. Thus, we must be able to name the service with an assertable identity and to grant rights to that identity without contradicting the governing local policy.
3. *Dynamic establishment of trust domains*. In order to coordinate resources, VOs need to establish trust among not only users and resources in the VO but also among the VO's resources, so that they can be coordinated. These trust domains can span multiple organizations and must adapt dynamically as participants join, are created, or leave the VO.

Traditional means of security administration that involve manual editing of policy databases or issuance of credentials cannot meet the demands of these dynamic scenarios. We require a user-driven security model that allows users to create entities and policy domains in order to create and coordinate resources within VOs.

## 3 GT2 Grid Security Model

To set the stage for our discussion of OGSA and GT3 security, we review briefly the security technologies incorporated in the Globus Toolkit version 2 (GT2) [13] and explain how these technologies address the three key issues introduced above. GT2 includes services for Grid Resource Allocation and Management (GRAM), Monitoring and Discovery (MDS), and data movement (GridFTP). These services use a common Grid Security Infrastructure (GSI) [4, 15] to provide security functionality.

*Diverse site security mechanisms*. GSI defines a common credential format based on X.509 identity certificates [5, 28] and a common protocol based on transport layer security (TLS [10], SSL [18]). An X.509 certificate, in conjunction with an associated private key, forms a unique credential set that a Grid entity (service or user) uses to authenticate itself to other Grid entities; the TLS-based protocol is used to perform authentication and then provide message protection (encryption, integrity checking), as desired, on the subsequent data stream. Gateways are used to translate between this common GSI infrastructure and local site mechanisms. For example, the Kerberos Certificate Authority (KCA) [29] and SSLK5/PKINIT provide translation from Kerberos to GSI and vice versa, respectively. These mechanisms allow a site with an existing Kerberos infrastructure to continue using that installation and convert credentials between Kerberos and GSI as needed.

Each GSI certificate is issued by a trusted party known as a certificate authority (CA), usually run by a large organization or commercial company. In order to trust the X.509 certificate presented by an entity, one must trust the CA that issued the certificate. We chose to use X.509 identity certificates within GSI because establishment of this trust is relatively lightweight. In contrast to mechanisms such as Kerberos [25], where inter-institutional trust requires a bilateral agreement at the organizational level, trust in a CA can be established unilaterally: A single entity in an organization can decide to trust any CA, without necessarily involving the organization as a whole. This feature is key to the establishment of VOs that involve only some portion of an organization, where the organization as a whole may provide little or no support for the VO.

*Dynamic creation of entities and the granting of privileges to those entities*. GSI introduces X.509 proxy certificates, a GSI extension to X.509 identity certificates [28] that allow a user to assign dynamically a new X.509 identity to an entity and then delegate some subset of their rights to that identity. Users create a proxy certificate by issuing a new X.509 certificate signed using their own credentials instead of involving a CA. This mechanism allows new credentials and identities to be created quickly without the involvement of a traditional administrator.

*Dynamic creation and management of overlaid trust domains*. The requirement for overlaid trust domains to establish VOs is satisfied by GSI using both proxy certificates and security services such as the Community Authorization Service (CAS) [26]. GSI has an implicit policy that any two entities bearing proxy certificates issued by the same user will inherently trust each other. This policy allows users to create trust domains dynamically by issuing proxy certificates to any services that they wish to have collaborate.

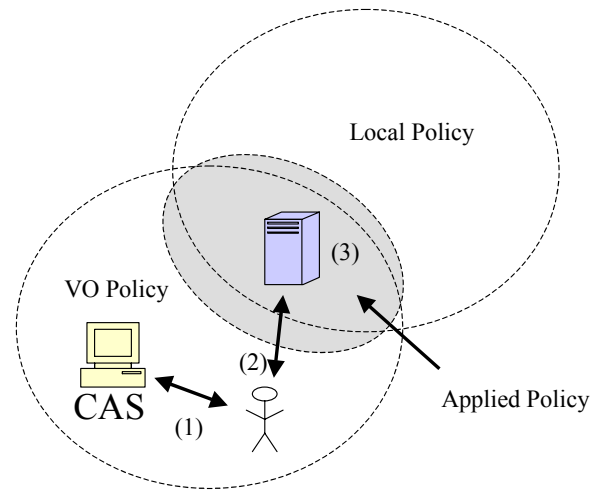

**Figure 2. CAS allows VOs to express policy, and it allows resources to apply policy that is a subset of VO and local policy.**

The implicit policy of trust between proxy holders allows for the lightweight, dynamic creation of simple trust domains but has limitations when it comes to complicated trust domains with, for example, limited trust between multiple parties. In that latter case, we can turn to security services such as CAS that allow flexible, expressive policy to be created regarding multiple users in a VO. CAS allows a VO to express the policy that has been outsourced to it by the resource providers in the VO. This process, as illustrated in Figure 2, comprises three steps:
1. The user authenticates to CAS and receives assertions from CAS expressing the VO's policy in terms of how that user may use VO resources.
2. The user then presents the assertion to a VO resource along with the usage request.
3. In evaluating whether to allow the request, the resource checks both local policy and the VO policy expressed in the CAS assertion.

CAS allows a resource to remain the ultimate authority over that resource, but it also allows the VO to control a subset of that enforced policy. In turn, the VO can coordinate the policy across a number of resources to control the sharing of those resources by the VO.

In designing GSI we evaluated several related efforts before electing to build on PKI. We noted the following shortcomings in other approaches with respect to Grid security requirements:

- Kerberos [25] requires the explicit involvement of site administrators to establish interdomain trust relationships or to create new entities.
- The CRISIS wide area security system [3] defines a uniform and scalable security infrastructure for wide area systems but does not address interoperability with local security mechanisms.
- Secure SHell (SSH) [30] provides a strong system of authentication and message protection but has no support for translation between different mechanisms or for creation of dynamic entities.
- The Legion security model [19] is perhaps the most similar to that of GT2, using X.509 certificates for delegation. However, it lacks mechanisms for creation of dynamic entities.

## 4 The GT3 Security Model for OGSA

We now turn to the question of how Grid security challenges can be addressed within the context of the Open Grid Services Architecture (OGSA) [14], a set of technical specifications that align Grid technologies with emerging Web services technologies [18].

Web services technologies allow software components to be defined in terms of access methods, bindings of these methods to specific communication mechanisms, and mechanisms for discovering relevant services. While particular mechanisms and methods are not prescribed, some mechanisms are emerging as ubiquitous. In particular, the Simple Object Access Protocol (SOAP) [1] provides a means of messaging using XML envelopes to encapsulate payloads, with HTTP the most commonly used underlying protocol. Another example is the Web Services Description Language (WSDL) [7], which provides a method for expressing operation signatures and bindings to protocols and endpoints in an XML document (groups of operations bundled together to form a "Web service").

OGSA defines standard Web service interfaces and behaviors that add to Web services the concepts of stateful services and secure invocation, as well as other capabilities needed to address Grid-specific requirements that are not relevant for this paper. These interfaces and behaviors define what is called a "Grid service" and allow users to manage the Grid service's life-cycle, as allowed by policy, and to create sophisticated distributed services. Grid services can define, as part of their interface, service data elements (SDEs) that other entities can (again, subject to policy) query or subscribe to.

OGSA introduces both new opportunities and new challenges for Grid security. Emerging Web services security specifications address the expression of Web service security policy (WS-Policy [2], XACML [33]), standard formats for security token exchange (WS-Security [22], SAML [32]), and standard methods for authentication and establishment of security contexts and trust relationships (WS-SecureConversation [20], WS-Trust [21]). These specifications may be exploited to create standard, interoperable methods for these features in Grid security. But they may, in some cases, also need to be extended to address the Grid security requirements listed above.

Version 3 of the Globus Toolkit (GT3) and its accompanying Grid Security Infrastructure (GSI3) provide the first implementation of OGSA mechanisms. GT3's security model seeks to allow applications and users to operate on the Grid in as seamless and automated a manner as possible. Security mechanisms should not have to be instantiated in an application but instead should be supplied by the surrounding Grid infrastructure, allowing the infrastructure to adapt on behalf of the application to meet the application's requirements. The application should need to deal with only application-specific policy. GT3 uses the following powerful features of OGSA and Web services security to work toward this goal:

1. Casts security functionality as OGSA services to allow them to be located and used as needed by applications.
2. Uses sophisticated *hosting environments* to handle security for applications and allow security to adapt without having to change the application.
3. Publishes service security policy so that clients can discover dynamically what credentials and mechanisms are needed to establish trust with the service.
4. Specifies standards for the exchange of security tokens to allow for interoperability.

In the following subsections we describe how each of these features is used in our GT3 OGSA security model. We then explain how they are used together to support seamless Grid security. We emphasize that what we describe here is our own GT3 OGSA security model, that is, an OGSA-based security model that we (the Globus Project security team) have defined and implemented within GT3. Ultimately, we hope that an appropriate standards body (e.g., the Global Grid Forum) will define *the* OGSA security model, which may or may not incorporate some of the ideas presented here.

## 4.1 Security as Services

Secure operation in a Grid environment requires that applications and services be capable of supporting a variety of security functionality, such as authentication, authorization, credential conversion, auditing, and delegation. Grid applications need to interact with other applications and services that have a range of security mechanisms and requirements. These mechanisms and requirements are likely to evolve over time as new mechanisms are developed or policies change. Grid applications must avoid embedding security mechanisms statically in order to adapt to changing requirements.

Our OGSA security model casts security functions as OGSA services. This strategy allows well-defined protocols and interfaces to be defined for these services and permits an application to outsource security functionality by using a security service with a particular implementation to fit its current need.

A draft OGSA Security Roadmap [31] presented in 2002 to the Global Grid Forum itemizes numerous security services, including the following.

- *Credential processing service:* A service that handles the details of processing and validating authentication tokens
- *Authorization service:* A service that evaluates policy rules regarding the decision to allow the attempted actions based on information about the requestor (identity, attributes, etc.), the target (identity, policy, attributes, etc.), and details of the request.
- *Credential Conversion service:* A service that enables bridging of different trust or mechanism domains by converting credentials between trust roots or mechanisms.
- *Identity Mapping service:* A service that takes a user's identity in one domain and returns the identity in another (e.g., given the user's X.509 identity, it could return the Kerberos principal name).
- *Audit:* A service that securely logs relevant information about events.

## 4.2 Hosting Environment

It is not a trivial task to find and use security services such as those described in the preceding section: in fact, it can require considerable sophistication on the part of the application. Ideally, application developers should not be burdened with the details of this process.

Grid services, like the Web services they leverage, may be built on sophisticated container-based *hosting environments* such as J2EE or .Net. These hosting environments provide a high level of functionality and allow for much security implementation complexity to be pulled from applications. We envision that most security functionality will be placed in hosting environments, simplifying application development and allowing security functionality to be upgraded independently of applications.

## 4.3 Publishing of Security Policy

In order to establish trust, two entities need to be able to find a common set of security mechanisms that both understand. The use of hosting environments and security services, as described previously, enables OGSA applications and services to adapt dynamically and use different security mechanisms. However, an application can select the proper security mechanisms and credentials only if it knows what mechanisms and credentials are acceptable to the service with which it wishes to interact.

The WS-Policy [2] specification and its related specifications define how a Web service can publish its security policy along with its interface specification as part of a WSDL document. Such a published policy can express requirements for mechanisms, acceptable trust roots, token formats, and other security parameters. An application wishing to interact with the service can examine this published policy and gather the needed credentials and functionality by contacting appropriate OGSA security services.

## 4.4 Specified Format for Security Tokens

The WS-Security [22], WS-SecureConversation [20], and WS-Trust [21] specifications contain conventions and formats for the communication of various mechanism-specific tokens (e.g., Kerberos tickets and X.509 certificates) inside SOAP envelopes. This enveloping standardizes the protocol for security mechanisms and allows mechanisms to be independent of any application protocol. Hosting environments can recognize security-related messages and route them to an appropriate service for handling, and entities in the network can recognize whether and how an interaction is secured. For example, a firewall can recognize whether a connection is authenticated and allow only authenticated connections.

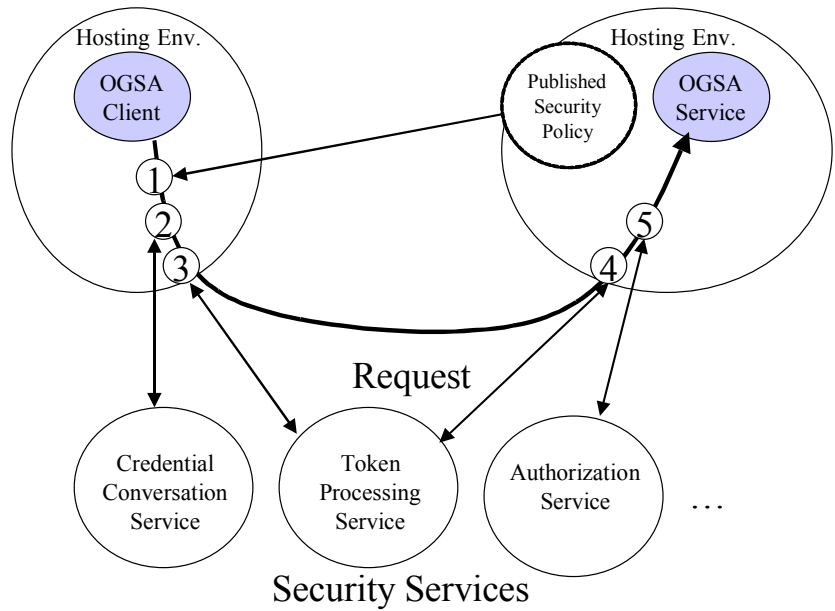

**Figure 3. Example of a secured request in the OGSA security model. The steps are described in the text.**

### 4.5 The GT3 OGSA Security Model in Action

Figure 3 shows a simplified example of the GT3 OGSA security model in action. A request is made by the OGSA client on the left to the OGSA service on the right. Both client and service are contained in advanced hosting environments (Section 4.2) that handle all security functionality for their respective contained application and service. For simplicity many details of the security process, such as auditing, client-side authorization, and privacy are omitted. These functions would be implemented similarly, with the hosting environments using OGSA services to provide the needed functionality.

The client first forms a request intended for the OGSA service and passes the request to its hosting environment for processing and delivery. The following steps are taken to handle the security of the request:

1. The client's hosting environment retrieves and inspects the security policy of the target service to determine what mechanisms and credentials are required to submit a request.
2. If the client's hosting environment determines that the needed credentials are not already present, it contacts a credential conversion service to convert existing credentials to the needed format, mechanism, and/or trust root. Two examples of such services are CAS [26], for translating the user's personal credential to a VO credential, and KCA [29], for converting between Kerberos and PKI mechanisms.
3. The client's hosting environment uses a token processing and validation service (e.g., XKMS [11]) to handle the formatting and processing of authentication tokens for exchange with the target service. This service relieves the application and its hosting environment from having to understand the details of any particular mechanism.
4. On the server side, the hosting environment likewise uses a token processing service to process the authentication tokens presented by the client. (In the example, both use the same service, but each could use a separate service.)
5. After authentication and the determination of client identity and attributes, the target service's hosting environment presents the details of the request and client information to an authorization service (e.g., PERMIS [6] or Akenti [27]) for a policy decision.

If all these steps complete successfully, the target service's hosting environment then presents the authorized request to the target service application. The application, knowing that the hosting environment has already taken care of security, can focus on application-specific request processing steps.

## 5 GT3 Security Implementation

The Grid Security Infrastructure version 3 (GSI3) of the Globus Toolkit version 3 is an initial implementation of key components of the OGSA security model described in

Section 4. This implementation has two key advantages over its GT2 predecessor described in Section 3:

- *Use of WS-Security protocols and standards.* GT3 uses SOAP and the Web services security specifications for all of its communications. This allows it to leverage and use standard current and future Web service tools and software.
- *Tight least-privilege model.* In contrast to GT2, the GT3 resource management implementation uses *no* privileged services. All privileged code is contained in two small, tightly constrained setuid programs.

We describe here how these two advantages are implemented in GT3 and describes the GT3 Grid Resource Acquisition and Management (GRAM) system, which illustrates all key GSI3 components.

## 5.1 Use of Web Services Security and Protocol

GT3 uses Web services specifications to allow security messages and secured messages to be transported, understood, and manipulated by standard Web services tools and software. GT3 offers both stateful and stateless forms of secured communication.

*Stateful.* Like GT2, GSI3 supports the establishment of a security context that serves to authenticate two parties to each other and allows for the exchange of secured messages between the two parties. GT2 uses the TLS [10] transport protocol for both security context establishment and message protection. Our GT3 implementation achieves security context establishment by implementing WS-SecureConversation [20] and WS-Trust [21], which use SOAP messages to transport context-establishment tokens between the two parties. The GT3 messages carry the same context establishment tokens used by GT2 but transports them over SOAP instead of TCP. Once the security context is established, GSI3 implements message protection using the Web services standards for secured messages (XML-Signature [34] and XML-Encryption [35]).

*Stateless.* To allow for communication without the initial establishment of a security context, GT3 also offers the ability to sign messages independent of any established security context, by using the XML-Signature specification. Thus, a message can be created and signed, allowing the recipient to verify the message's origin and integrity, without establishing synchronous communication with the recipient. A feature of this approach is that the identity of the recipient does not have to be known to the sender when the message is sent. As we discuss later, this allows for messages to be created by clients and delivered to services whose creation is caused by the message itself.

## 5.2 Tight Least-Privilege Model

"Least privilege" is a well-known principle in computer security that states that each entity should only have the minimal privilege needed to accomplish its assigned role and no more. GT3 introduces two notable features that improve its security through the least privilege principle.

*No privileged services.* Network services, since they accept and process communications from outside the resource, are prone to compromise by remote users through logic errors, buffer overflows, and the like. GT3 removes all privileges from these services, significantly reducing the impact of compromises by minimizing the privileges gained.

*Minimal privileged code.* In GT3, the privileged code is confined to small programs, each of which performs a specific function and works only with local users, accepting no network connections. The simple and well-constrained functionality of these programs allows them to be audited effectively and reduces the chance that they can be used maliciously to gain privilege elevation.

## 5.3 GT3 GRAM Implementation

We introduce the GSI3 implementation by describing how it is used in GT3's GRAM [8, 9] system. GRAM is a fundamental GT service enabling remote clients to instantiate, manage and monitor, in a secure fashion, computational tasks ("jobs") on remote resources. While GT3 offers a number of other services (e.g., for file movement and monitoring), GRAM is the most complicated service in GT3 from a security perspective because it provides for the secure, remote, dynamic instantiation of processes, involving both secured interaction with a remote client and the local operating system.

To invoke a job using GRAM, a client describes the job to be run, specifying such details as the name of the executable, the working directory, where input and output should be stored, and the queue in which it should run. This description is sent to the resource and ultimately results in the creation of an instance of a Managed Job Service (MJS). A MJS is a Grid service that acts as an interface to its associated job, instantiating it and then allowing it to be controlled and monitored with standard Grid and Web service mechanisms.

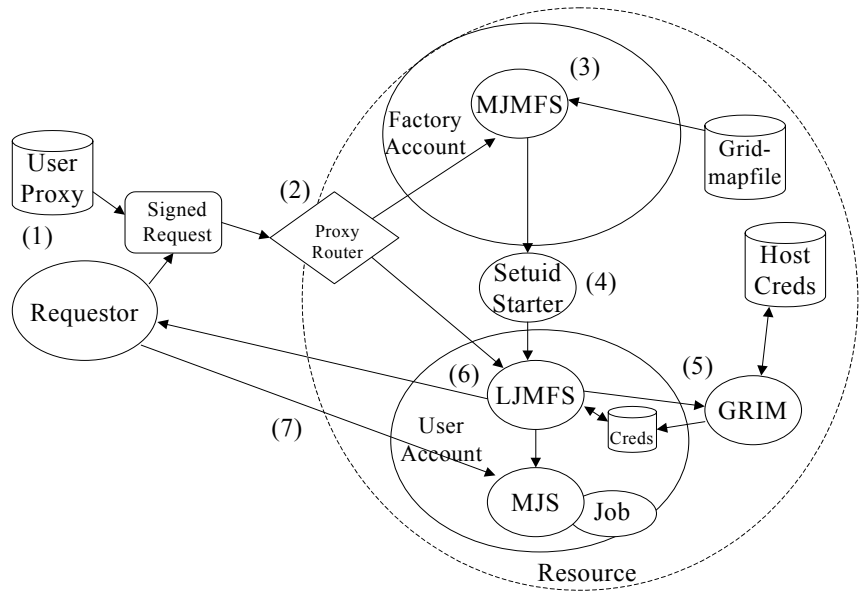

**Figure 4. A requestor initiating a job with the GT3 GRAM system. The steps are described in the text.**

An MJS is created by invoking a create operation on a MJS factory service. While conceptually we want to run one MJS factory service per user account, this approach is not ideal in practice because it involves resource consumption by factories that sit idle when the user is not using the resource. Thus, we introduce an additional construct, the Master Managed Job Factory Service (MMJFS). One MMJFS runs on each resource, in a non-privileged account, and invokes Local Managed Job Factory Services (LMJFS) for users in their account as needed. A service called a Proxy Router routes incoming requests from a user to either that user's LMJFS, if present, or the MMJFS, if a LMJFS is not present for the user making the request.

All MJS and MJS factories are implemented as Grid services running in a hosting environment. Each active account has a hosting environment running for its use, with a MJS factory and one or more MJS instances running in that hosting environment. This approach allows for the creation of multiple services in a lightweight manner.

Figure 4 shows a requestor initiating a job in the GT3 GRAM architecture. On the left is the requestor with a set of GSI user proxy credentials. The resource, with its GRAM services and host credentials, is on the right. Job initiation proceeds as follows.

1. The requestor forms a job description and signs it with appropriate GSI credentials. This request is sent to the target resource on which process initiation is desired.
2. The Proxy Router service accepts the request and either routes it to an LMJFS, if present (skip to step 6), or to the MMJFS otherwise (on to step 3).
3. The MMJFS verifies the signature on the request and establishes the identity of the requestor. It then determines the local account in which the job should be run based on the requestor's identity using the *grid-mapfile*, a local configuration file containing mappings from GSI identities to local identities [4].
4. The MMJFS invokes the *Setuid Starter* process to start a LMJFS for the requestor. The *Setuid Starter* is a privileged program (typically setuid-root) whose sole function is to start a pre-configured LMJFS for a user.
5. When a LMJFS starts, it needs to acquire credentials and register itself with the Proxy Router. To register, the LMJFS sends a message (not shown) to the Proxy Router. This informs the Proxy Router of the existence of the LMJFS so that it can route future requests for job initiation to it. The LMJFS invokes the Grid Resource Identity Mapper (*GRIM*) to acquire a set of credentials. *GRIM* is a privileged program (typically setuid-root) that accesses the local host credentials and from them generates a set of GSI proxy credentials for the LMJFS. This proxy credential has embedded in it the user's Grid identity, local account name, and local policy to help the requestor verify that the LMJFS is appropriate for its needs.
6. The LMJFS receives the signed job request. The LMJFS verifies the signature on the request to make sure it has not been tampered with and to verify the requestor is authorized to access the local user account in which the LMJFS is running. Once these verifications are complete, the LMJFS invokes an MJS with the job initiation request and returns the service reference of the MJS to the user.

7. The requestor connects to the MJS to initiate the job. The requestor and MJS perform mutual authentication, the MJS using the credentials acquired from *GRIM*. The MJS verifies that the requestor is authorized to initiate processes in the local account. The requestor authorizes the MJS as having a GRIM credential issued from an appropriate host credential and containing a Grid identity matching its own. This approach allows the client to verify that the MJS it is talking to is running not only on the right host but also in an appropriate account. When making this connection, the user also delegates GSI credentials to the MJS for the job.

## 5.4 Implementation Status

The architecture and features described in Section 5 are available as part of GT3 [36]. Some of the Web services specifications are still in the standards process, and we expect the details of our implementation to change to track these specifications as they progress

## 6 Conclusions

Grid computing presents a number of security challenges that are met by the Globus Toolkit's Grid Security Infrastructure (GSI). Version 3 of the Globus Toolkit (GT3) implements the emerging Open Grid Services Architecture; its GSI implementation (GSI3) takes advantage of this evolution to improve on the security model used in earlier versions of the toolkit. GSI3 remains compatible (in terms of credential formats) with those used in GT2, while eliminating privileged network services and making other improvements. Its development provides a basis for a variety of future work. In particular, we are interested in exploiting WS-Routing to improve firewall compatibility; in defining and implementing standard services for authorization, credential conversation, identity mapping; and in using WS-Policy to automate application determination of requirements and location of services that meet those requirements.

## Acknowledgments

We are pleased to acknowledge contributions to the GT3 security design and GSI3 implementation by Doug Engert, Thomas Sandholm, and Rachana Ananthakrishnan. We also thank Ruth Aydt and our anonymous reviewers for their helpful comments. This work was supported in part by the Mathematical, Information, and Computational Sciences Division subprogram of the Office of Advanced Scientific Computing Research, U.S. Department of Energy, under contracts W-31-109-Eng-38, DE-AC03-76SF0098, DE-FC03-99ER25397 and No. 53-4540-0080; and by IBM.